\documentclass[fleqn,10pt]{wlscirep}
\usepackage[utf8]{inputenc}
\usepackage[T1]{fontenc}

\usepackage{graphicx}%
\usepackage{multirow}%
\usepackage{amsmath,amssymb,amsfonts}%
\usepackage{amsthm}%
\usepackage{mathrsfs}%
\usepackage[title]{appendix}%
\usepackage{xcolor}%
\usepackage{textcomp}%
\usepackage{manyfoot}%
\usepackage{booktabs}%
\usepackage{algorithm}%
\usepackage{algorithmicx}%
\usepackage{algpseudocode}%
\usepackage{listings}%
\title{The Debugging Decay Index: Rethinking Debugging Strategies for Code LLMs}

\author[1,*]{Muntasir Adnan}
\author[1]{Carlos C. N. Kuhn}
\affil[1]{Open Source Institute, University of Canberra, Bruce, Canberra, Australia}

\affil[*]{Corresponding Author: Adnan.adnan@canberra.edu.au}


\begin{abstract}
The effectiveness of AI debugging follows a predictable exponential decay pattern; most models lose 60-80\% of their debugging capability within just 2-3 attempts, despite iterative debugging being a critical capability for practical code generation systems.
We introduce the Debugging Decay Index (DDI), a mathematical framework that quantifies when debugging becomes ineffective and predicts intervention points. Our strategic fresh start approach shifts from exploitation to exploration at strategic points in the debugging process, demonstrating that well-timed interventions can rescue the effectiveness of debugging. DDI reveals a fundamental limitation in current AI debugging and provides the first quantitative framework for optimising iterative code generation strategies.
\end{abstract}
\begin{document}

\flushbottom
\maketitle
%
%


\section*{Introduction}
\label{intro}
The advent of large language models (LLMs) has transformed automated code generation, enabling developers to produce functional code with remarkable speed and scale~\cite{jiang2024surveylargelanguagemodels}. 
Recent efforts have shifted toward debugging-based code generation, where LLMs iteratively refine their output based on compiler feedback or error messages, mirroring traditional software development practices~\cite{ldb, mapcoder, pycapsule, teaching_debugging}. 
This iterative approach represents a fundamental departure from single-pass generation, yet the underlying dynamics of debugging-based LLM-guided code generation remain critically underexplored.
Existing studies often apply an arbitrary number of debugging attempts without examining their optimal extent or effectiveness over continuous iterations~\cite{ldb, mapcoder, pycapsule}. 
This approach incurs significant computational costs and lacks methodological rigour in determining when additional iterations cease to yield meaningful improvements.
Preliminary research and our analysis suggest that LLM-guided debugging typically follows an exponential decay pattern, where debugging effectiveness diminishes rapidly with successive attempts~\cite{pycapsule}. 
However, no systematic work has been conducted to characterise this decay phenomenon or explore strategies to break these patterns for improved performance.
This pattern of diminishing returns in iterative LLM approaches extends beyond code generation, with recent research on reasoning models demonstrating similar complexity-dependent limitations where self-correction capabilities plateau and models either overthink simple problems or fail entirely on complex ones~\cite{shojaee2025illusion}, suggesting a natural ceiling that warrants systematic investigation.

Furthermore, as debugging-based LLM-guided code generation becomes increasingly prevalent, evaluation metrics must evolve beyond traditional single-pass assessments~\cite{passk, passk_review} to account for the iterative nature of the process.
Current evaluation approaches treat code as static artefacts rather than as the product of a dynamic development process, overlooking the significant quality enhancements that often emerge through systematic debugging and refinement~\cite{debugging_tech}.
This limitation becomes increasingly problematic as the field moves toward debugging-based approaches that more closely align with human software development practices~\cite{effective_debugging}.
Single-pass metrics such as $pass@k$~\cite{passk} measure the probability that at least one correct solution exists among $k$ independently generated candidates.
However, this approach does not guarantee diversity among the generated candidates. 
It fails to account for the iterative debugging process that is central to practical software development workflows~\cite{effective_debugging}, and relies solely on manually written test cases~\cite{codejudge}.

This study examines the effectiveness of repeated debugging attempts in LLM-based code generation and investigates strategic interventions to enhance the debugging process. 
To address the limitations of existing evaluation metrics, we propose a novel evaluation framework: the \textbf{Debugging Decay Index (DDI)}. 
The DDI metric provides a unified assessment of LLM coding proficiency by modelling the exponential effectiveness decay observed in iterative debugging processes.
Our framework computes strategic intervention timing $t_\theta$ based on configurable effectiveness decay thresholds $\theta$, returning a comprehensive evaluation tuple $(E_0, \lambda, t_\theta, R^2)$ that captures initial performance, decay sustainability, strategic stopping points, and model fit quality. 
This multi-dimensional approach enables nuanced evaluation across different aspects of the code generation and debugging pipeline.
Our investigation addresses the following research questions:

\begin{itemize}
    \item \textbf{RQ1 (Debugging Window)}: How many debugging attempts maximise the effectiveness of LLM-generated code before further iterations yield diminishing returns, and how do these attempt windows vary across different model architectures?

    \item \textbf{RQ2 (DDI)}: How can we develop a unified evaluation metric that comprehensively assesses LLM code generation and debugging capabilities, quantifying initial performance, sustained effectiveness, and iterative refinement capability encompassing both reasoning proficiency and instruction-following competency across diverse model architectures?

    \item \textbf{RQ3 (Strategic Fresh Starts)}: Based on the optimal debugging windows identified in RQ1 and the decay characteristics quantified in RQ2, to what extent can implementing fresh start strategies (reinitiating the code generation process) after reaching effectiveness thresholds improve overall accuracy compared to continued iterative refinement within the same generation context?
\end{itemize}

\section*{Literature Review}
\label{lit}
\subsection*{Evaluation Metric}
Code-generating LLMs are typically evaluated based on functional correctness or whether the generated code effectively solves the given task.
In this paradigm, the $ pass@k$ metric~\cite {passk} has become a standard measure.
Pass@k is the probability that at least one of $k$ independently generated solutions to a problem passes all unit tests. 
Pass@k can be written as -
\begin{center}
    $pass@k = 1 - \mathbb{P}(\textit{all incorrect})$
\end{center}
The unbiased~\cite{passk, passk_review} estimation formula is - 
\begin{center}
    $pass@k = 1 - \frac{\binom{n-c}{k}}{\binom{n}{k}}$
\end{center}
Where $n$ is the total number of samples generated, $n\geq k$ and $c$ of them pass.
One can draw $n\ge k$ samples and count the number of solutions $c$ that pass~\cite{passk_review}.  
Numerous subsequent works on LLM-guided code generation have used $pass@k$. 
For example, CodeT~\cite{codet} and Top Pass~\cite {top_pass} evaluated various models on standard benchmarks using the $pass@k$ metric. 
In MBR-EXEC~\cite{natural_lang_to_code}, authors measured pass@k for HumanEval~\cite{passk}, Mostly Basic Python Programming(MBPP)~\cite {mbpp} to compare instruction tuning. 
Code generation benchmark leaderboards and evaluations of programming-focused large language models consistently report pass@k metrics (typically k=1, 5, 10, and occasionally up to k=100) as a standard method for model comparison~\cite{alphacode, gpt4, qwen25coder, evalperf, evalplus}.
The elegance of this metric lies in its simplicity and direct correlation with functionality; a model that can generate at least one correct solution within k attempts demonstrates meaningful capability in code generation tasks.
Importantly, $pass@k$ is a binary, functional metric; it only cares whether any generated solution is entirely correct.

Building upon this foundation, researchers have conducted thorough investigations into the pass@k metric's characteristics, examining its sensitivity to both the sample size ($k$) and the inherent difficulty of programming problems~\cite{alphacode, a_survey_eval, evalplus, wang2025llmsreplacehumanevaluators}.
A critical limitation identified is the metric's sole reliance on provided test suites, which may not comprehensively verify all aspects of code correctness or efficiency~\cite{a_survey_eval}.
This concern was empirically validated when researchers augmented the standard HumanEval benchmark with more rigorous test cases (creating HumanEval+~\cite{evalplus}), resulting in a significant performance drop of approximately 20-30\% across various models. 
A more fundamental concern relates to how optimising for $pass@k$ can distort model behaviour and evaluation priorities.
Top Pass~\cite{top_pass} introduced a ranking model that directly optimises for this metric, revealing a key limitation: $pass@k$ rewards getting one solution correct over producing multiple near-correct solutions. 
This approach fails to reward quick convergence and may allow models to game the metric by generating variants of the same algorithm rather than exploring diverse approaches.
Complementary findings revealed that 42\% of code generations failing unit tests were still rated valuable by programmers and proposed a hybrid metric~\cite{align_passk_sim} combining functional correctness with syntactic similarity, which achieved a 14\% stronger correlation with programmer-perceived value. 
These findings suggest that evaluation metrics should consider not only binary correctness but also how effectively code can be refined through debugging.
In response to these limitations, several research works have proposed several variations of pass@k. 
The count@k metric~\cite{coderujb} counts how many of k attempts are correct, while AlphaCode introduced n@k~\cite{alphacode} that generalises pass@k to measure exactly n correct solutions out of k attempts. 
Addressing the need to recognise partially correct solutions, the pass-ratio@n metric~\cite{passratio} averages the squared test-pass ratio across n generated code samples. 
This approach gives partial credit to nearly-correct solutions, addressing the granularity that $pass@k$ lacks.

While these functionality-based metrics dominate code generation evaluation, many researchers still report non-functional metrics such as BLEU~\cite{bleu}, CodeBLEU~\cite{codebleu}, or ROUGE~\cite{rouge} to measure syntactic similarity. 
These metrics are not replacements for $pass@k$ but often accompany it to gauge quality aspects beyond functional correctness. 
While a few orthogonal approaches exist, they all fail to capture the iterative nature of code development and the debugging capabilities of LLMs.

Our proposed Debugging Decay Index (DDI) addresses this gap by focusing on the iterative path to functional correctness rather than arbitrary sampling.
Unlike traditional metrics, 
DDI measures how effectively models leverage debugging feedback to improve a solution until it achieves functional correctness iteratively.
This approach acknowledges that real-world programming rarely involves generating multiple independent attempts; instead, developers iteratively refine their code through debugging cycles.
By quantifying the efficiency of this debugging process, DDI provides a reliable evaluation of how models would perform in practical software development contexts, where strategic iteration, rather than random sampling, is the path to successful code.

\subsection*{Debugging}
Researchers have explored dynamic approaches to incorporate execution feedback and debugging capabilities in LLM-guided code generation. 
Recent work~\cite{revisitselfdebuggingselfgeneratedtests} investigated debugging in two distinct contexts: in-context debugging, which involves inspecting intermediate execution states, and post-context debugging, which focuses on analysing error results after complete execution. 
Building on this foundation, the SELF-DEBUGGING framework~\cite{teaching_debugging} demonstrated how LLMs can analyse execution results and explain their own generated code line by line, mirroring approaches developed initially for human developers~\cite{thomas2019pragmatic}.
The framework allowed for a maximum of 10 debugging attempts, but the researchers observed that successful debugging typically concluded within just three iterations.
By comparison, MapCoder~\cite{mapcoder} implemented a more extensive debugging protocol, allowing up to 25 attempts, but limiting them to a maximum of 5 attempts per individual plan. 
The authors reported that while increased debugging iterations generally improved performance, this relationship was not strictly linear across all datasets. 
Notably, their results for HumanEval-ET did not follow the expected proportional improvement trend, indicating potential dataset-specific considerations in debugging efficacy.
Similarly, the Large Language Model Debugger (LDB)~\cite{ldb} employed 10 debugging attempts in their standard configuration, with additional experiments using up to 20 attempts on the HumanEval dataset. 
Their findings revealed a continuous but diminishing improvement trend, with gains becoming increasingly marginal after the fifth attempt.
The subsequent 15 attempts collectively yielded only 2.4\% additional improvement. 
PyCapsule~\cite{pycapsule} implemented a more streamlined approach compared to MapCoder while still achieving state-of-the-art (SOTA) performance across several benchmark datasets. 
The framework employed just five debugging attempts beyond the initial solution and fitted the normalised debugging effectiveness to an exponential decay function, revealing that effectiveness usually diminishes dramatically after the third attempt and follows an exponential decay pattern.
Their analysis further demonstrated that debugging effectiveness varies significantly across model architectures: OpenAI's GPT-4~\cite{gpt4} exhibited complete loss of debugging effectiveness (relative to the first attempt) by the third iteration, while GPT-3.5~\cite{gpt4} showed similar exhaustion by the fourth attempt. 
In contrast, Qwen2.5-coder-instruct~\cite{qwen25coder} maintained some debugging capability until the fifth attempt, suggesting model-specific patterns in debugging performance decay.
These findings highlight a critical research gap: the need for a standardised approach to quantify and optimise debugging iterations for LLM code generation. 

Empirical evidence across debugging frameworks reveals consistent diminishing returns, though the specific decay characteristics vary systematically across model architectures, suggesting model-specific debugging signatures that remain unexplored as evaluation criteria. 
Existing approaches treat these decay patterns as inevitable limitations rather than quantifiable characteristics of the model. 
This systematic variation in debugging persistence presents an opportunity to develop methodologies that both measure debugging capability through decay modelling (RQ2) and identify possible optimal intervention strategies when effectiveness diminishes (RQ3).

\section*{Methodology}\label{method}
\subsection*{\textbf{RQ1:} Debugging Window}
We introduce the concept of a ``debugging window" in the context of LLMs for code generation, which refers to the threshold for debugging attempts.
While diminishing effectiveness will always occur with continued debugging efforts, establishing this window allows us to determine a practical cutoff point that balances debugging effectiveness with computational efficiency.
To model the effectiveness of each debugging attempt over time, this study employs the exponential decay function (Equation~\ref{eq:exp_decay_func}).
The exponential decay function is defined as follows:
\begin{equation}
    E(t) = E_0 e^{-\lambda t} \label{eq:exp_decay_func}
\end{equation}
In this study, $E(t)$ represents the effectiveness of debugging at attempt $t$, while $E_0$ denotes the initial effectiveness corresponding to the very first attempt. 
The decay constant $\lambda$ represents the rate of effectiveness loss over successive attempts and serves as our primary metric for benchmarking model performance.
Models with lower $\lambda$ values maintain their effectiveness longer across debugging iterations, and $t$ represents the discrete number of debugging attempts, allowing us to model the temporal progression of debugging effectiveness.
To further analyse the decay process, we examine the half-life $t_{1/2}$, which represents the number of debugging attempts after which the effectiveness reduces to half its initial value $E_0$. 
By definition and from Equation~\ref{eq:exp_decay_func}, we get:
\begin{equation}
    E(t_{1/2}) = \frac{1}{2} E_0 
    \implies t_{1/2} = \frac{\ln(2)}{\lambda} \label{eq:half_life_def}
\end{equation}
We can generalise Equation~\ref{eq:half_life_def} to determine the number of debugging attempts required for any given decay percentage. 
For a decay threshold where effectiveness can lose up to $\theta\%$ of its initial value (meaning $(100-\theta)\%$ effectiveness remains), the number of debugging attempts $t_\theta$ is given by:
\begin{equation}
    t_\theta = \frac{\ln\left(\frac{100}{100-\theta}\right)}{\lambda} \label{eq:ttheta_formula}
\end{equation}
This generalised formula enables us to calculate the debugging window for any threshold $\theta$, providing the flexibility to determine when diminishing effectiveness justifies terminating the debugging process based on specific computational constraints.

\subsection*{RQ2: The Debugging Decay Index (DDI)}
Our proposed DDI integrates our exponential decay analysis from RQ1 
to create a comprehensive evaluation framework for LLM debugging capabilities. 
Unlike traditional metrics that focus solely on final outcomes, DDI captures the efficiency and capability of the debugging process and the final accuracy.

\subsubsection*{Mathematical Formulation}
The DDI is formulated as a function 
\begin{equation*}
    DDI(data, \theta) \rightarrow (E_0, \lambda, t_\theta, R^2)
\end{equation*} 
that accepts \textit{data}, the normalised debugging effectiveness measurements across multiple iterative attempts; and $\theta$, the effectiveness decay threshold(s) representing the maximum acceptable performance degradation. 
Following the PyCapsule~\cite{pycapsule} framework, the normalised debugging effectiveness data represents the independent influence of each debugging attempt.
The DDI framework identifies strategic intervention points $t_\theta$ where debugging effectiveness would degrade by $\theta\%$ from the initial value. 
In RQ3, we leverage these DDI-calculated intervention points to evaluate whether implementing fresh start strategies at the predicted timing can improve overall accuracy compared to continued iterative refinement within the same generation context. 
Fresh starts involve reinitiating the debugging process with the original problem statement only.
DDI returns a four-element tuple:
\begin{itemize}
    \item \textbf{$E_0$ (Initial Effectiveness)}: $E_0$ represents the initial effectiveness, calculated as $E_0 = N_{solved\_at\_attempt\_0} / N_{total}$. 
    This metric is directly comparable to $pass@1$ and represents the model's inherent code generation capability before any debugging.

    \item \textbf{$\lambda$ (Decay Rate)}: The decay constant extracted from fitting the exponential decay function (Equation~\ref{eq:exp_decay_func}) to normalised debugging effectiveness data. 
    A lower $\lambda$ indicates slower decay in effectiveness and more persistent debugging behaviour, reflecting sustained instruction following and reasoning consistency across iterations.

    
    
    \item \textbf{$t_\theta$ (Optimal Intervention Points)}: $t_\theta$ represents the maximum number of debugging attempts before effectiveness drops by $\theta\%$ from the initial value.
    This represents the strategic intervention threshold corresponding to the $\theta$ value, calculated using Equation~\ref{eq:ttheta_formula}.
    Since debugging attempts must be discrete integers, we apply the ceiling function to convert the continuous mathematical solutions into practical stopping points.
    This ensures that the debugging window provides sufficient attempts to reach at least the specified effectiveness threshold. 

    \item \textbf{$R^2$ (Fit Quality)}: The coefficient of determination measuring how well the exponential decay model explains the observed debugging effectiveness patterns. 
    We interpret the results using the following categories: Excellent ($R^2 \geq 0.9$), Good ($0.7 \leq R^2 < 0.9$), or Poor ($R^2 < 0.7$). 
    High $R^2$ values indicate predictable exponential decay behaviour, while low values suggest erratic or non-exponential debugging patterns that may require alternative evaluation approaches.

\end{itemize}

\subsubsection*{Evaluation Process and Interpretation}
The DDI evaluation proceeds through four core steps: initial assessment records $E_0 = N_{solved\_at\_0} / N_{total}$; iterative debugging tracks effectiveness 
at each attempt; decay analysis fits Equation~\ref{eq:exp_decay_func} using nonlinear least squares regression to extract $\lambda$, setting $\lambda = \text{None}$ when insufficient data points ($n < 3$) exist; and threshold calculation determines strategic intervention timing $t_\theta$ using Equation~\ref{eq:ttheta_formula}.

The DDI outputs provide comprehensive model characterisation of code generation and debugging capabilities, requiring interpretation of both effectiveness metrics and fit quality.
For models with high $R^2$ values ($\geq 0.7$), the combination of $E_0$ and $\lambda$ reveals distinct model archetypes: high $E_0$ + low $\lambda$ indicates both strong reasoning and persistent debugging (ideal), low $E_0$ + low $\lambda$ suggests consistent but ineffective approaches, high $E_0$ + high $\lambda$ indicates strong initial reasoning but poor debugging persistence, while low $E_0$ + high $\lambda$ represents both weak reasoning and rapid debugging degradation.
However, for models with poor fit quality ($R^2 < 0.7$),
the exponential decay assumption may not apply, indicating that a different mathematical function may be required to fully characterise the model behaviour. 
In such cases,
evaluation should rely primarily on $E_0$ when using DDI.
Pseudocode for DDI is provided in Appendix: DDI Pseudocode.

\subsection*{RQ3: Strategic Fresh Starts}
\label{subsec: fresh}

To investigate whether strategic interventions can mitigate the debugging decay phenomenon identified in RQ2, we implement fresh start strategies at DDI calculated strategic intervention points. 
A fresh start completely clears conversation history and begins anew with only the original problem statement.
This mechanism addresses the rapid degradation of effectiveness observed in the exponential decay pattern, particularly when models become trapped in the low-effectiveness tail, where continued debugging attempts yield negligible improvement. 
The fresh start strategy operates on the hypothesis that reinitialising the generation process shifts the model from exploiting failing solution approaches back to exploring alternative solution spaces.
Based on empirical evidence from RQ2, we observe varied suitable intervention points $t_\theta$ across different models, as demonstrated in Table~\ref{tab: ddi_results_table}. 
Given the variance in decay patterns, we strategically implement fresh starts at DDI-calculated intervention thresholds, enabling each model to benefit from reinitialisation at its optimal timing.
To ensure a fair comparison with existing approaches, we maintain the same total attempt budget as previous works~\cite{pycapsule, mapcoder}, consisting of six attempts (initial generation plus five debugging iterations). 
Our approach strategically allocates these attempts while triggering fresh starts at DDI-calculated intervention points, testing whether strategic reinitialisation can overcome debugging decay while maintaining strict comparability with baseline methods.

\section*{Evaluation and Experimental Setup}
To address our research questions regarding debugging windows (RQ1) and the DDI framework (RQ2), we applied our calculation methodology to eighteen SOTA language models using the HumanEval~\cite{passk} dataset. 
Our experimental design systematically evaluates the decay patterns of debugging effectiveness across diverse model architectures, ranging from smaller, specialised models like DeepSeek-Coder 6.7b~\cite{guo2024deepseek} to larger, general-purpose models such as Claude-3-7-sonnet-20250219~\cite{claude}, GPT-4~\cite{gpt4}, and GPT-3.5~\cite{gpt4}.
Using normalised debugging effectiveness data from HumanEval~\cite{passk}, we extracted model-specific decay constants $\lambda$. 
For each model, we calculated $E_0, \lambda, t_\theta \text{ where } \theta \in 50, 80, 90, 95, 99 \text{ and } R^2$.
Additionally, we report $A_0$ values representing the final accuracy achieved after six attempts without any fresh start interventions (same as PyCapule~\cite{pycapsule}), providing a baseline performance metric for comparison with our strategic restart approaches in RQ3, see Table~\ref{tab:model_performance}.

\begin{figure}[!h]
    \centering
    \includegraphics[width=\linewidth]{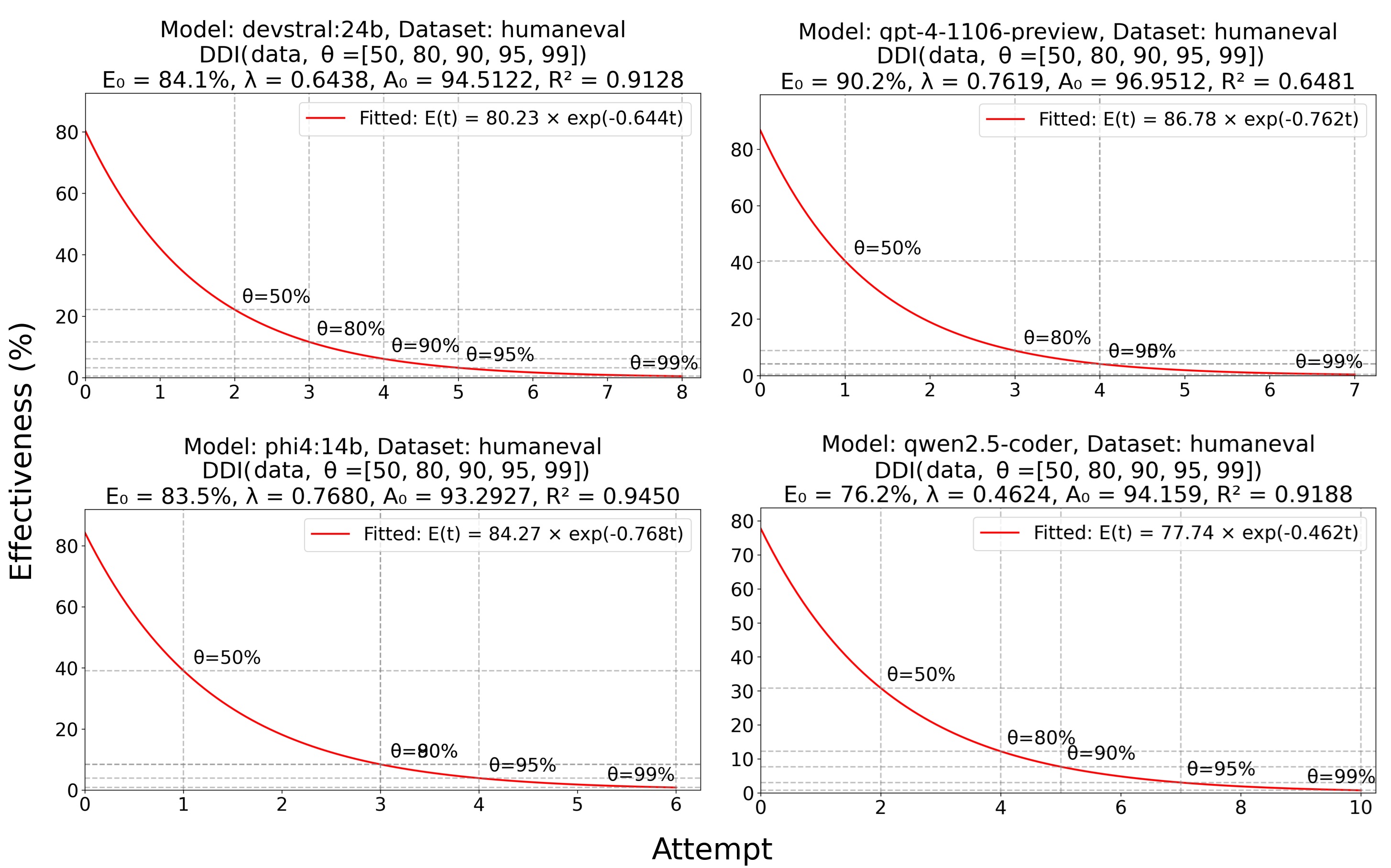}
    \caption{Exponential decay curves fitted to debugging effectiveness data for four language models. 
    The grey dashed lines indicate effectiveness thresholds at different $\theta$ values. 
    The $\lambda$ (decay rate) and $R^2$ (goodness-of-fit) values are displayed for each model.
    }
    \label{fig:decay_percentages}
\end{figure}

Table~\ref{tab: ddi_results_table} and Figure~\ref{fig:decay_percentages} present our comprehensive analysis of debugging decay characteristics across these LLMs. 
The debugging window calculations reveal distinct performance characteristics across model architectures. 

\begin{table}[!t]
  
    \centering
    \begin{tabular}{p{4.2cm}p{1cm}p{1cm}p{0.8cm}p{2.2cm}p{1cm}}
        \hline
        \textbf{Model} & $E_0$ & \textbf{$\lambda$} & \textbf{$A_0$} & \textbf{$t_\theta$} & $R^2$\\
        \hline
        \textbf{claude-3-7-sonnet-20250219} & 93.902 & None & 100.00 & [] & None\\
        \textbf{codegemma:7b} & 51.219 & 0.9309 & 66.463 & [1, 2, 3, 4, 5] & Excellent\\
        \textbf{codellama:7b} & 21.341 & 0.2467 & 45.122 & [3, 7, 10, 13, 19] & Poor\\
        \textbf{codestral:22b} & 58.537 & 0.3388 & 89.024 & [3, 5, 7, 9, 14] & Good\\
        \textbf{deepseek-coder-v2:16b} & 71.951 & 0.9692 & 84.146 & [1, 2, 3, 4, 5] & Excellent\\
        \textbf{deepseek-coder:6.7b} & 45.732 & 0.4737 & 74.390 & [2, 4, 5, 7, 10] & Excellent\\
        \textbf{devstral:24b} & 84.146 & 0.6438 & 94.512 & [2, 3, 4, 5, 8] & Excellent\\
        \textbf{gemma2:9b} & 59.146 & 0.7632 & 76.219 & [1, 3, 4, 4, 7] & Excellent\\
        \textbf{gpt-3.5-turbo} & 73.781 & 1.3297 & 82.317 & [1, 2, 2, 3, 4] & Excellent\\
        \textbf{gpt-3.5-turbo-1106} & 70.732 & 0.7553 & 85.976 & [1, 3, 4, 4, 7] & Excellent\\
        \textbf{gpt-4-1106-preview} & 90.244 & 0.7619 & 96.951 & [1, 3, 4, 4, 7] & Poor\\
        \textbf{granite3.3:8b} & 68.902 & 0.9482 & 82.317 & [1, 2, 3, 4, 5] & Excellent\\
        \textbf{llama2:7b} & 3.659 & 0.1185 & 10.976 & [6, 14, 20, 26, 39] & Poor\\
        \textbf{llama3.1:8b} & 56.707 & 1.1142 & 72.561 & [1, 2, 3, 3, 5] & Excellent\\
        \textbf{mistral:instruct} & 29.878 & 0.5291 & 54.268 & [2, 4, 5, 6, 9] & Excellent\\
        \textbf{phi4-reasoning:14b} & 59.146 & 0.6052 & 81.098 & [2, 3, 4, 5, 8] & Excellent\\
        \textbf{phi4:14b} & 83.537 & 0.7680 & 93.293 & [1, 3, 3, 4, 6] & Excellent\\
        \textbf{qwen2.5-coder} & 76.219 & 0.4624 & 94.159 & [2, 4, 5, 7, 10] & Excellent\\
        \hline
    \end{tabular}
    \caption{DDI Results for Different Models for $\theta \in \{50, 80, 90, 95, 99\}$ on the HumanEval dataset. 
    $R^2$ indicates exponential fit quality: Excellent ($R^2 \geq 0.9$), Good ($0.7 \leq R^2 < 0.9$), Poor ($R^2 < 0.7$). 
    Models with $\lambda = \text{None}$ had insufficient data points for exponential fitting after filtering zero effectiveness values.}
    \label{tab: ddi_results_table}
\end{table} 

Claude-3.7-Sonnet demonstrated remarkable performance, achieving 100\% effectiveness ($A_0 = 100\%$) essentially within two attempts, which prevented fitting to the exponential decay model, resulting $\lambda=None$. 
This exceptional performance represents a unique case where conventional debugging window calculations may not apply.

Conversely, the Phi-4~\cite{abdin2024phi} model comparison provides particularly revealing insights into the relationship between reasoning capabilities and debugging sustainability. 
While phi4:14b~\cite{abdin2024phi} $(E_0=83.537\%, \lambda=0.76)$ significantly outperformed phi4-reasoning:14b~\cite{abdin2024phi} $(E_0=59.146\%, \lambda=0.60)$ in initial effectiveness by approximately 24\%, likely due to phi4-reasoning not being instruction fine-tuned and thus more challenging to parse, the reasoning model demonstrated remarkable debugging improvement capacity.
Despite starting from a substantially lower baseline, phi4-reasoning achieved a final accuracy of 81.098\% compared to phi4:14b's 93.293\%, representing an improvement of 21.95\% versus only 9.75\%, respectively. 
The reasoning model improved more than twice as much as the standard model through iterative debugging.
These findings suggest that the decay constant $\lambda$ captures not only debugging efficiency but also underlying reasoning capabilities and instruction adaptability. 
The reasoning model's lower $\lambda$ value (0.591 vs 0.711) indicates superior debugging sustainability, enabling it to extract more value from iterative refinement processes. 
This reveals that reasoning-capable models, although potentially harder to prompt initially, possess an enhanced capacity for systematic error correction and solution refinement —a crucial characteristic for extended debugging sessions where sustained improvement matters more than initial performance.

GPT variants exhibit relatively fast effectiveness decay, with gpt-3.5-turbo~\cite{gpt4} showing the highest decay rate, reaching the 80\% threshold by attempts 2-3. 
In contrast, models like codestral:22b~\cite{codestral} and deepseek-coder:6.7b~\cite{guo2024deepseek} demonstrate more sustained debugging capabilities with lower decay rates ($\lambda=0.375 \text{ and } \lambda=0.330$ respectively), extending debugging windows to 5-7 attempts for the same threshold.
DDI reveals nuanced debugging characteristics that would be missed by simple effectiveness metrics alone. 
The case of phi4-reasoning:14b~\cite{abdin2024phi} exemplifies this.

To evaluate the effectiveness of strategic fresh starts proposed in RQ3, we implemented restart interventions at the calculated strategic thresholds for $\theta \in \{50, 80\}$ effectiveness degradation. 
Table~\ref{tab:model_performance} presents the comparative performance results, demonstrating the impact of strategic reinitialisation versus continued iterative debugging.
The results reveal that strategic fresh starts can significantly improve debugging performance across most models without requiring any additional computational resources. 
Since fresh starts only involve clearing conversation history at predetermined intervention points while maintaining the same attempt budget, the computational overhead remains equivalent with similar or reduced token usage on average compared to continuous debugging sessions. 
For example, DeepSeek-Coder-V2-16B reduced token consumption from 108,289 to approximately 89,000 tokens on average, while Codestral-22B maintained usage around 94,000 tokens compared to 97,000 in the continuous sessions. 

\begin{figure}
    \centering
    \includegraphics[width=\linewidth]{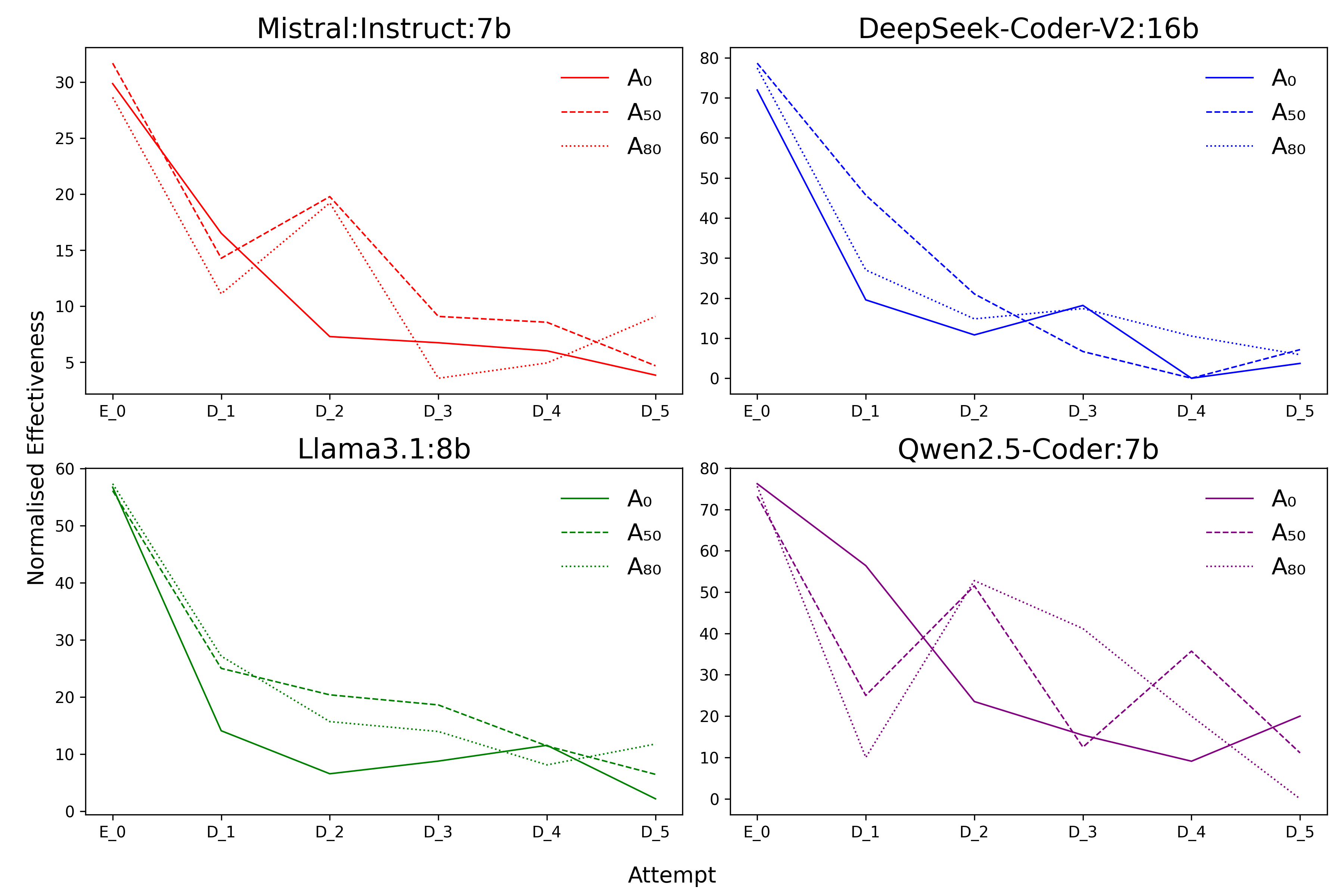}
    \caption{Normalised debugging effectiveness trajectories compared to baseline continuous debugging ($A_0$) with fresh start strategies implemented at $\theta \in \{50, 80\}$. 
    The distinctive spikes in $A_{50} \text{ and } A_{80}$ demonstrate successful intervention effects, where fresh starts reset the debugging process and break the monotonic decay pattern observed in baseline approaches. 
    These spikes represent moments where strategic reinitialisation successfully shifts models from failed solution exploitation back to productive exploration, enabling recovery from debugging decay within the same computational budget.}
    \label{fig:break_exp}
\end{figure}

Of the six models evaluated, all showed performance improvements when fresh starts were applied at  DDI-calculated intervention points. 
Significantly, llama3.1:8b~\cite{touvron2023llama} showed the most significant improvement, enhancing its baseline accuracy from 72.56\% to 82.82\%. 
In contrast, deepseek-coder-v2:16b~\cite{guo2024deepseek} experienced the second largest enhancement, with its baseline accuracy increasing from 84.1\% to 92.1\%.
Similarly, Mistral:Instruct~\cite{jiang2023mistral7b} demonstrated consistent gains across both thresholds, improving from 54.3\% to 62.8\% and 57.3\%.
This demonstrates that strategic timing of fresh starts, rather than simply increasing attempt counts, can overcome debugging decay patterns and improve overall effectiveness.
Analysis of the normalised debugging effectiveness patterns (Figure~\ref{fig:break_exp}) reveals that fresh start interventions successfully break the exponential decay curve observed in RQ1. 
Rather than following the predicted decay trajectory, models implementing fresh starts at strategic intervention points demonstrate renewed effectiveness spikes, essentially resetting the decay pattern and enabling continued productive debugging. 
This empirical evidence supports our hypothesis that strategic reinitialisation shifts models from exploitation of failing solution approaches back to exploration of alternative solution spaces.

\begin{table}[h]
  
    \centering
    \begin{tabular}{p{3cm}p{2.5cm}p{2.5cm}p{2.5cm}}
        \hline
        \textbf{Model} & \textbf{$A_0$} & \textbf{$A_{50}$} & \textbf{$A_{80}$} \\
        \hline
        codegemma:7b & 66.4634 & \textbf{71.9512} & 70.1220 \\
        codestral:22b & 89.0244 & 88.4146 & \textbf{91.4634} \\
        deepseek-coder-v2:16b & 84.1463 & \textbf{92.0732} & 90.2439 \\
        devstral:24b & 94.5122 & 93.9024 & \textbf{95.1220} \\
        granite3.3:8b & 82.3171 & \textbf{86.5854} & 85.3659 \\
        llama3.1:8b & 72.5610 & \textbf{82.3171} & 81.7073 \\
        mistral:instruct & 54.2683 & \textbf{62.8049} & 57.3171 \\
        phi4:14b & 93.2927 & 93.9024 & \textbf{96.3415} \\
        qwen2.5-coder & 94.1588 & \textbf{95.1220} & 94.5122 \\
        \hline
    \end{tabular}
    \caption{Performance comparison showing baseline accuracy $A_0$ achieved within six attempts without intervention, versus fresh start strategies implemented at DDI-calculated intervention points where $\theta \in \{50, 80\}$. 
    $A_{50}$ and $A_{80}$ represent final accuracy when fresh starts are triggered at $t_{50}$ and $t_{80}$ thresholds respectively. 
    The corresponding intervention timing ($t_\theta$ values) for each model can be found in Table~\ref{tab: ddi_results_table}. 
    Bold values indicate performance improvements over the baseline $A_0$, demonstrating cases where strategic reinitialisation outperforms continued iterative debugging within the same debugging context at no extra token usage at all}
    \label{tab:model_performance}
\end{table}

\section*{Limitations and Future Work}


\subsection*{Generalisation Limitations}
The primary limitation of DDI lies in its dataset-specific nature. 
While exponential decay patterns appear robust across multiple debugging contexts~\cite{pycapsule, ldb}, specific $\lambda$ and $t_\theta$ values depend on problem set characteristics, including complexity distribution and debugging challenge types. 
Like all evaluation metrics, DDI parameters require dataset-specific calibration and cannot be directly compared across different problem sets without statistical validation.
Effective generalisation requires either statistically sufficient numbers of diverse datasets to establish robust parameter ranges or standardised benchmarks with comprehensive complexity coverage that can serve as universal debugging evaluation foundations. 
Due to computational constraints, our current validation focuses on HumanEval variants, though the underlying exponential decay phenomenon has been observed across broader programming contexts including other benchmark datasets~\cite{pycapsule}.
Additionally, while our fresh start interventions demonstrate performance improvements across almost all evaluated models, the magnitude of these improvements critically depends on the selected effectiveness threshold $\theta$. 
Although we observe consistent benefits regardless of threshold selection, selecting $\theta$ values for maximum performance gains represents a crucial but unexplored aspect of our framework. 
The systematic selection of strategic intervention thresholds falls outside the scope of this study and represents an important direction for future investigation.

\subsection*{Research Directions}
The DDI framework opens several promising research directions. 
Cross-dataset validation across diverse programming benchmarks would establish parameter ranges and calibration requirements for broader applicability. 
A critical immediate direction involves developing principled methods for optimal threshold selection, potentially through adaptive strategies that learn from model-specific decay characteristics or problem complexity indicators.

Perhaps most significantly, comparative analysis of human versus AI debugging decay patterns could validate the theoretical foundations of exponential effectiveness degradation, potentially revealing fundamental principles of iterative problem-solving that extend beyond code generation to broader cognitive tasks. 
The mathematical framework's simplicity and interpretability make it well-suited for such interdisciplinary investigation, offering a quantitative foundation for understanding debugging effectiveness across both artificial and biological intelligence systems. 
Importantly, the flexibility of the framework extends beyond modelling exponential decay. 
By applying the same underlying concept, we can incorporate various functions that represent the decay patterns observed in the input data, allowing us to extract the decay rate specific to the model in question.
While we employed exponential functions, as they provided the best fit for the majority of evaluated models, the DDI methodology can readily incorporate linear decay, polynomial regression, or other mathematical functions, depending on the observed model behaviour. 
This adaptability ensures the framework remains robust across different model architectures and debugging contexts, where alternative decay patterns may emerge as the field evolves.

\section*{Conclusion}
This work introduces the Debugging Decay Index (DDI), a novel evaluation framework that characterises the exponential effectiveness decay patterns inherent in LLM-guided iterative debugging processes. 
Through a systematic analysis of eighteen state-of-the-art (SOTA) language models on HumanEval, we demonstrate that debugging effectiveness typically follows predictable exponential decay trajectories, enabling a principled determination of optimal intervention timing rather than relying on arbitrary attempt limits.
Our key contributions include the mathematical characterisation of debugging decay patterns across diverse model architectures; the DDI framework, which provides a unified assessment of the coding and debugging abilities of LLMs; and the demonstration that strategic fresh start interventions at DDI-calculated thresholds can break exponential decay patterns and improve final accuracy without incurring additional computational costs.
The fresh start strategy proves particularly effective, requiring zero additional computational resources, and demonstrates that strategic intervention, rather than increased attempt budgets, drives performance improvements. 
The DDI framework provides immediate practical value for researchers seeking to optimise debugging workflows, while also opening up promising research directions in adaptive intervention strategies and cross-dataset validation. 
\section*{Additional Information}
The authors declare no competing interests.

\section*{Data Availability}
All benchmark datasets utilised in this study are openly available from their respective public repositories. 
The code and analysis scripts will be made publicly available upon publication.

\bibliography{sample}

\begin{thebibliography}{10}
\urlstyle{rm}
\expandafter\ifx\csname url\endcsname\relax
  \def\url#1{\texttt{#1}}\fi
\expandafter\ifx\csname urlprefix\endcsname\relax\def\urlprefix{URL }\fi
\expandafter\ifx\csname doiprefix\endcsname\relax\def\doiprefix{DOI: }\fi
\providecommand{\bibinfo}[2]{#2}
\providecommand{\eprint}[2][]{\url{#2}}

\bibitem{jiang2024surveylargelanguagemodels}
\bibinfo{author}{Jiang, J.}, \bibinfo{author}{Wang, F.}, \bibinfo{author}{Shen, J.}, \bibinfo{author}{Kim, S.} \& \bibinfo{author}{Kim, S.}
\newblock \bibinfo{title}{A survey on large language models for code generation} (\bibinfo{year}{2024}).
\newblock \eprint{2406.00515}.

\bibitem{ldb}
\bibinfo{author}{Zhong, L.}, \bibinfo{author}{Wang, Z.} \& \bibinfo{author}{Shang, J.}
\newblock \bibinfo{title}{Debug like a human: A large language model debugger via verifying runtime execution step by step}.
\newblock In \bibinfo{editor}{Ku, L.-W.}, \bibinfo{editor}{Martins, A.} \& \bibinfo{editor}{Srikumar, V.} (eds.) \emph{\bibinfo{booktitle}{Findings of the Association for Computational Linguistics: ACL 2024}}, \bibinfo{pages}{851--870} (\bibinfo{publisher}{Association for Computational Linguistics}, \bibinfo{address}{Bangkok, Thailand}, \bibinfo{year}{2024}).

\bibitem{mapcoder}
\bibinfo{author}{Islam, M.~A.}, \bibinfo{author}{Ali, M.~E.} \& \bibinfo{author}{Parvez, M.~R.}
\newblock \bibinfo{title}{{M}ap{C}oder: Multi-agent code generation for competitive problem solving}.
\newblock In \emph{\bibinfo{booktitle}{Proceedings of the 62nd Annual Meeting of the Association for Computational Linguistics (Volume 1: Long Papers)}}, \bibinfo{pages}{4912--4944} (\bibinfo{publisher}{Association for Computational Linguistics}, \bibinfo{address}{Bangkok, Thailand}, \bibinfo{year}{2024}).

\bibitem{pycapsule}
\bibinfo{author}{Adnan, M.}, \bibinfo{author}{Xu, Z.} \& \bibinfo{author}{Kuhn, C. C.~N.}
\newblock \bibinfo{title}{Large language model guided self-debugging code generation} (\bibinfo{year}{2025}).
\newblock \eprint{2502.02928}.

\bibitem{teaching_debugging}
\bibinfo{author}{Chen, X.}, \bibinfo{author}{Lin, M.}, \bibinfo{author}{Sch{\"a}rli, N.} \& \bibinfo{author}{Zhou, D.}
\newblock \bibinfo{title}{Teaching large language models to self-debug}.
\newblock In \emph{\bibinfo{booktitle}{The 12th International Conference on Learning Representations}} (\bibinfo{year}{2024}).

\bibitem{shojaee2025illusion}
\bibinfo{author}{Shojaee, P.} \emph{et~al.}
\newblock \bibinfo{journal}{\bibinfo{title}{The illusion of thinking: Understanding the strengths and limitations of reasoning models via the lens of problem complexity}}.
\newblock {\emph{\JournalTitle{arXiv preprint arXiv:2506.06941}}}  (\bibinfo{year}{2025}).

\bibitem{passk}
\bibinfo{author}{Chen, M.} \emph{et~al.}
\newblock \bibinfo{title}{Evaluating large language models trained on code} (\bibinfo{year}{2021}).
\newblock \eprint{2107.03374}.

\bibitem{passk_review}
\bibinfo{author}{Paul, D.~G.}, \bibinfo{author}{Zhu, H.} \& \bibinfo{author}{Bayley, I.}
\newblock \bibinfo{journal}{\bibinfo{title}{Benchmarks and metrics for evaluations of code generation: A critical review}}.
\newblock {\emph{\JournalTitle{IEEE AITest}}}  (\bibinfo{year}{2024}).

\bibitem{debugging_tech}
\bibinfo{author}{Liu, A.} \& \bibinfo{author}{Coblenz, M.}
\newblock \bibinfo{title}{Debugging techniques in professional programming}.
\newblock In \emph{\bibinfo{booktitle}{13th Annual Workshop at the Intersection of PL and HCI}} (\bibinfo{year}{2023}).

\bibitem{effective_debugging}
\bibinfo{author}{Spinellis, D.}
\newblock \emph{\bibinfo{title}{Effective Debugging: 66 Specific Ways to Debug Software and System}} (\bibinfo{publisher}{Addison-Wesley Professional}, \bibinfo{address}{Boston, MA}, \bibinfo{year}{2016}).

\bibitem{codejudge}
\bibinfo{author}{Tong, W.} \& \bibinfo{author}{Zhang, T.}
\newblock \bibinfo{title}{Codejudge: Evaluating code generation with large language models}.
\newblock In \emph{\bibinfo{booktitle}{Proceedings of the 2024 Conference on Empirical Methods in Natural Language Processing}} (\bibinfo{year}{2024}).

\bibitem{codet}
\bibinfo{author}{Chen, B.} \emph{et~al.}
\newblock \bibinfo{journal}{\bibinfo{title}{Codet: Code generation with generated tests}}.
\newblock {\emph{\JournalTitle{arXiv preprint arXiv:2207.10397}}}  (\bibinfo{year}{2022}).

\bibitem{top_pass}
\bibinfo{author}{Lyu, Z.-C.}, \bibinfo{author}{Li, X.-Y.}, \bibinfo{author}{Xie, Z.} \& \bibinfo{author}{Li, M.}
\newblock \bibinfo{journal}{\bibinfo{title}{Top pass: Improve code generation by pass@k-maximized code ranking}}.
\newblock {\emph{\JournalTitle{Frontiers of Computer Science}}}  (\bibinfo{year}{2024}).

\bibitem{natural_lang_to_code}
\bibinfo{author}{Shi, F.}, \bibinfo{author}{Fried, D.}, \bibinfo{author}{Ghazvininejad, M.}, \bibinfo{author}{Zettlemoyer, L.} \& \bibinfo{author}{Wang, S.~I.}
\newblock \bibinfo{title}{Natural language to code translation with execution}.
\newblock In \emph{\bibinfo{booktitle}{Proceedings of the 2022 Conference on Empirical Methods in Natural Language Processing}} (\bibinfo{year}{2022}).

\bibitem{mbpp}
\bibinfo{author}{Austin, J.} \emph{et~al.}
\newblock \bibinfo{journal}{\bibinfo{title}{Program synthesis with large language models}}.
\newblock {\emph{\JournalTitle{arXiv preprint arXiv:2108.07732}}}  (\bibinfo{year}{2021}).

\bibitem{alphacode}
\bibinfo{author}{Li, Y.} \emph{et~al.}
\newblock \bibinfo{journal}{\bibinfo{title}{Competition-level code generation with alphacode}}.
\newblock {\emph{\JournalTitle{Science}}} \textbf{\bibinfo{volume}{378}}, \bibinfo{pages}{1092--1097} (\bibinfo{year}{2022}).

\bibitem{gpt4}
\bibinfo{author}{{OpenAI}} \emph{et~al.}
\newblock \bibinfo{title}{Gpt-4 technical report} (\bibinfo{year}{2024}).
\newblock \eprint{2303.08774}.

\bibitem{qwen25coder}
\bibinfo{author}{Hui, B.} \emph{et~al.}
\newblock \bibinfo{title}{Qwen2.5-coder technical report} (\bibinfo{year}{2024}).
\newblock \eprint{2409.12186}.

\bibitem{evalperf}
\bibinfo{author}{Liu, J.} \emph{et~al.}
\newblock \bibinfo{title}{Evaluating language models for efficient code generation}.
\newblock In \emph{\bibinfo{booktitle}{First Conference on Language Modeling}} (\bibinfo{year}{2024}).

\bibitem{evalplus}
\bibinfo{author}{Liu, J.}, \bibinfo{author}{Xia, C.~S.}, \bibinfo{author}{Wang, Y.} \& \bibinfo{author}{Zhang, L.}
\newblock \bibinfo{title}{Is your code generated by chat{GPT} really correct? rigorous evaluation of large language models for code generation}.
\newblock In \emph{\bibinfo{booktitle}{Thirty-seventh Conference on Neural Information Processing Systems}} (\bibinfo{year}{2023}).

\bibitem{a_survey_eval}
\bibinfo{author}{Chen, L.} \emph{et~al.}
\newblock \bibinfo{journal}{\bibinfo{title}{A survey on evaluating large language models in code generation tasks}}.
\newblock {\emph{\JournalTitle{arXiv preprint arXiv:2408.16498}}}  (\bibinfo{year}{2024}).

\bibitem{wang2025llmsreplacehumanevaluators}
\bibinfo{author}{Wang, R.} \emph{et~al.}
\newblock \bibinfo{title}{Can llms replace human evaluators? an empirical study of llm-as-a-judge in software engineering}, \doiprefix\url{10.1145/3728963} (\bibinfo{year}{2025}).
\newblock \eprint{2502.06193}.

\bibitem{align_passk_sim}
\bibinfo{author}{Dibia, V.} \emph{et~al.}
\newblock \bibinfo{title}{Aligning offline metrics and human judgments of value for code generation models}.
\newblock In \bibinfo{editor}{Rogers, A.}, \bibinfo{editor}{Boyd-Graber, J.} \& \bibinfo{editor}{Okazaki, N.} (eds.) \emph{\bibinfo{booktitle}{Findings of the Association for Computational Linguistics: ACL 2023}}, \bibinfo{pages}{8516--8528}, \doiprefix\url{10.18653/v1/2023.findings-acl.540} (\bibinfo{publisher}{Association for Computational Linguistics}, \bibinfo{address}{Toronto, Canada}, \bibinfo{year}{2023}).

\bibitem{coderujb}
\bibinfo{author}{Zeng, Z.}, \bibinfo{author}{Wang, Y.}, \bibinfo{author}{Xie, R.}, \bibinfo{author}{Ye, W.} \& \bibinfo{author}{Zhang, S.}
\newblock \bibinfo{title}{Coderujb: An executable and unified java benchmark for practical programming scenarios}.
\newblock In \emph{\bibinfo{booktitle}{Proceedings of the 33rd ACM SIGSOFT International Symposium on Software Testing and Analysis}} (\bibinfo{year}{2024}).

\bibitem{passratio}
\bibinfo{author}{Yeo, S.}, \bibinfo{author}{Ma, Y.-S.}, \bibinfo{author}{Kim, S.~C.}, \bibinfo{author}{Jun, H.} \& \bibinfo{author}{Kim, T.}
\newblock \bibinfo{journal}{\bibinfo{title}{Framework for evaluating code generation ability of large language models}}.
\newblock {\emph{\JournalTitle{ETRI Journal}}} \textbf{\bibinfo{volume}{46}}, \bibinfo{pages}{106--117}, \doiprefix\url{10.4218/etrij.2023-0357} (\bibinfo{year}{2024}).

\bibitem{bleu}
\bibinfo{author}{Papineni, K.}, \bibinfo{author}{Roukos, S.}, \bibinfo{author}{Ward, T.} \& \bibinfo{author}{Zhu, W.}
\newblock \bibinfo{title}{Bleu: a method for automatic evaluation of machine translation}.
\newblock In \emph{\bibinfo{booktitle}{Proceedings of the 40th Annual Meeting of the Association for Computational Linguistics}}, \bibinfo{pages}{311--318} (\bibinfo{publisher}{Association for Computational Linguistics}, \bibinfo{address}{Philadelphia, Pennsylvania, USA}, \bibinfo{year}{2002}).

\bibitem{codebleu}
\bibinfo{author}{Ren, S.} \emph{et~al.}
\newblock \bibinfo{title}{Codebleu: a method for automatic evaluation of code synthesis} (\bibinfo{year}{2020}).
\newblock \eprint{2009.10297}.

\bibitem{rouge}
\bibinfo{author}{Lin, C.-Y.}
\newblock \bibinfo{title}{Rouge: A package for automatic evaluation of summaries}.
\newblock In \emph{\bibinfo{booktitle}{Text Summarization Branches Out}}, \bibinfo{pages}{74--81} (\bibinfo{publisher}{Association for Computational Linguistics}, \bibinfo{address}{Barcelona, Spain}, \bibinfo{year}{2004}).

\bibitem{revisitselfdebuggingselfgeneratedtests}
\bibinfo{author}{Chen, X.} \emph{et~al.}
\newblock \bibinfo{title}{Revisit self-debugging with self-generated tests for code generation} (\bibinfo{year}{2025}).
\newblock \eprint{2501.12793}.

\bibitem{thomas2019pragmatic}
\bibinfo{author}{Thomas, D.} \& \bibinfo{author}{Hunt, A.}
\newblock \emph{\bibinfo{title}{The Pragmatic Programmer: Your Journey to Mastery, 20th Anniversary Edition}} (\bibinfo{publisher}{Pearson Education}, \bibinfo{address}{Boston, MA}, \bibinfo{year}{2019}), \bibinfo{edition}{20th anniversary edition} edn.

\bibitem{guo2024deepseek}
\bibinfo{author}{Guo, D.} \emph{et~al.}
\newblock \bibinfo{journal}{\bibinfo{title}{Deepseek-coder: When the large language model meets programming--the rise of code intelligence}}.
\newblock {\emph{\JournalTitle{arXiv preprint arXiv:2401.14196}}}  (\bibinfo{year}{2024}).

\bibitem{claude}
\bibinfo{author}{{Anthropic}}.
\newblock \bibinfo{title}{Claude 3.7 sonnet and claude code}.
\newblock \bibinfo{howpublished}{Online} (\bibinfo{year}{2025}).
\newblock \bibinfo{note}{Https://www.anthropic.com/news/claude-3-7-sonnet}.

\bibitem{abdin2024phi}
\bibinfo{author}{Abdin, M.} \emph{et~al.}
\newblock \bibinfo{journal}{\bibinfo{title}{Phi-4 technical report}}.
\newblock {\emph{\JournalTitle{arXiv preprint arXiv:2412.08905}}}  (\bibinfo{year}{2024}).

\bibitem{codestral}
\bibinfo{author}{{MistralAI}}.
\newblock \bibinfo{title}{Codestral}.
\newblock \bibinfo{howpublished}{Online} (\bibinfo{year}{2024}).
\newblock \bibinfo{note}{Https://mistral.ai/news/codestral}.

\bibitem{touvron2023llama}
\bibinfo{author}{Touvron, H.} \emph{et~al.}
\newblock \bibinfo{journal}{\bibinfo{title}{Llama: Open and efficient foundation language models}}.
\newblock {\emph{\JournalTitle{arXiv preprint arXiv:2302.13971}}}  (\bibinfo{year}{2023}).

\bibitem{jiang2023mistral7b}
\bibinfo{author}{Jiang, A.~Q.} \emph{et~al.}
\newblock \bibinfo{title}{Mistral 7b} (\bibinfo{year}{2023}).
\newblock \eprint{2310.06825}.

\end{thebibliography}

\section*{Author contributions}
Muntasir Adnan and Carlos C. N. Kuhn collaborated on exploring the idea, writing, designing the experiments and reviewing the manuscript.

\section*{Legends}
\begin{table}[!h]
\centering
\begin{tabular}{|l|l|l|}
\hline
Title & Reference & Description \\
\hline
Table 1 & Table~\ref{tab: ddi_results_table} & DDI Results \\
\hline
Figure 1 & Figure~\ref{fig:decay_percentages} & Fitted Exponential Decay Curve \\
\hline
Figure 2 & Figure~\ref{fig:break_exp} & Fresh Start Effect \\
\hline
Table 2  & Table~\ref{tab:model_performance} & Fresh Start Results \\
\hline
\end{tabular}
\end{table}


\end{document}